\documentstyle[12pt,amsfonts] {article}
\begin{document}
\begin{titlepage}
\title
{On   quantum mechanics  in Friedmann--Robertson--Walker universe }
\author{{\large E A Tagirov}\\
N N Bogoliubov Laboratory of Theoretical Physics,\\
Joint Institute for Nuclear Research, Dubna, 141980, Russia,\\
    e--mail: tagirov@thsun1.jinr.dubna.su}

\date{}
\end{titlepage}
\maketitle

\newcommand{\vp}{\varphi}
\newcommand{\1}{{\bf\hat 1}}
\newcommand{\Oc}{O\left(c^{-2(N+1)}\right)}
\newcommand{\im}{{\rm i}} 
\newcommand{\Sg}{\Sigma}
\newcommand{\bgt}{\bigotimes}
\newcommand{\ptl}{\partial}
\newcommand{\Sche}{ Schr\"odinger equation\ }
\newcommand{\Schr}{Schr\"odinger representation\ }
\newcommand{\eu}{$ E_{1,3} $}
\newcommand{\rif}{V_{1,3}}
\newcommand{\ri}{$V_{1,3}$ }
\newcommand{\ov}{\overline}
\newcommand{\stc}{\stackrel}
\newcommand{\defst}{\stackrel{def}{=}}
\newcommand{\h}{\hbar}
\newcommand{\beq}{\begin{equation}}
\newcommand{\nde}{\end{equation}}
\newcommand{\beqa}{\begin{eqnarray}}
\newcommand{\ndea}{\end{eqnarray}}
\newcommand{\opif}{ p_{(i)} }
\newcommand{\opkf}{ p_{(k)} }
\newcommand{\opjf}{ p_{(j)} }
\newcommand{\hpif}{{\hat p}_{(i)}}
\newcommand{\hpkf}{{\hat p}_{(k)}}
\newcommand{\hpjf}{{\hat p}_{(j)}}
\newcommand{\oqif}{q^{(i)}}
\newcommand{\oqjf}{q^{(j)}}
\newcommand{\oqkf}{q^{(k)}}
\newcommand{\hqif}{{\hat q}^{(i)}}
\newcommand{\hqjf}{{\hat q}^{(j)}}
\newcommand{\hqkf}{{\hat q}^{(k)}}
\newcommand{\opi}{$ p_{(i)}$ }
\newcommand{\opk}{$ p_{(k)}$ }
\newcommand{\opj}{$ p_{(j)}$ }
\newcommand{\hpi}{${\hat p}_{(i)}$}
\newcommand{\hpk}{${\hat p}_{(k)}$}
\newcommand{\hpj}{${\hat p}_{(j)}$}
\newcommand{\oqi}{$q^{(i)}$}
\newcommand{\oqj}{$q^{(j)}$}
\newcommand{\oqk}{$q^{(k)}$}
\newcommand{\hqi}{${\hat q}^{(i)}$}
\newcommand{\hqj}{${\hat q}^{(j)}$}
\newcommand{\hqk}{${\hat q}^{(k)}$}
\begin{abstract}
 It is shown  that only in the  space--times admitting a 1+3-foliation by
 flat hypersurfaces (i.e., in the Bianchi I type space-times,
the isotropic version of which is the spatially flat
Friedmann--Robertson--Walker space-times )
the canonical quantization of geodesic motion and the quantum mechanical
asymptotics of the quantum theory of  scalar field lead to
the same canonical
commutation relations (CCR). Otherwise, the field--theoretical approach
leads to a deformation of CCR (particularly, operators of coordinates
do not commute),  and the Principle of Correspondence is
broken in a sense.   Thus, the spatially  flat cosmology is
distinguished intrinsically in the quantum theory.
\end{abstract}

\newpage
In my  paper \cite{TAG99}, the  generally covariant
 quantum mechanics (QM) of a neutral
spinless point--like  particle  in {\it the  general
Riemannian space--time} \ri was  formulated  as a one-particle
approximation in the quantum theory of
scalar field (QFT) $\vp$. This field--theoretical (FT) approach
QM in \ri is  a natural  alternative  to
approches based  on  quantization of the correspomding classical
mechanics, that is the  geodesic dynamics. There are  different procedures
of quantization:
canonical (see for application in \ri , e.g., in \cite{OLIV89, TAGI99}
and references  therein),  paths integration (see \cite{OLIV89,
MAR95}),  quasiclassical and geometrical
(see, e.g., \cite{DEW57, DOW73, SNIA80}).   However,  it should be noted  
that, except   \cite{TAGI99, SNIA80}, the mentioned as well
as numerious other papers devoted to the "free motion" on  curved
background    consider only   the case
of $\rif\ \sim\  T\otimes V_3$ for which the geodesic lines in \ri and
 $V_3$ are in the  one-to-one correspondence. For a time-dependent
\ri, this restriction is equivalent to the  non-relativistic
approximation.

A natural quantum  hierarchy of the fundamental physical theories
can be  represented   by the following diagram (I use the term
`dequantization' to denote a transition which  is  opposite
to quantization):
{\mathsurround=0pt
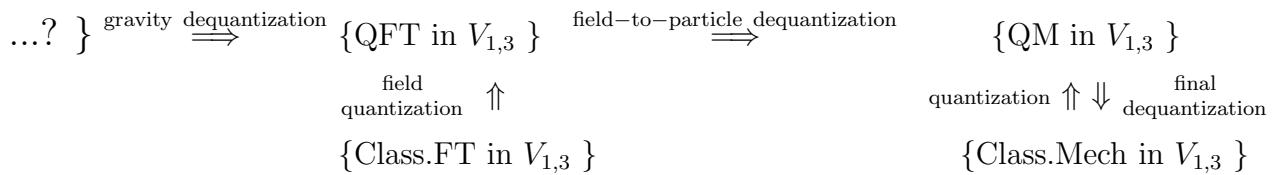
\begin{tabbing}
{\large ...? \}}  $\stc{\rm gravity\ dequantization}\Longrightarrow$
 \= \{QFT in \ri\}\=
\quad$\stc{\rm field-to-particle\ dequantization}{\Longrightarrow}$\quad\=
 \quad\quad \{QM in \ri\}  \\ [5pt]
\> ${{\rm field} \atop {\rm quantization}}\ \Uparrow$ \>
\>
{\scriptsize\rm quantization} $
\Uparrow\ \Downarrow {{\rm final}\atop {\rm dequantization}}$  \\[5pt]
  \> \{Class.FT in \ri\}  \> \>\quad \{Class.Mech in \ri\}
\end{tabbing}
}
\noindent
Do the procedures along the arrows starting from the classical
field theory and classical
mechanics lead to the same   QM,  at least, in the
simplest case of  spinless and chargeless particle and field?
They do  almost trivially in the case of the Minkowski space-time \eu
if the Cartesian coordinates are used. However, the situation  changes
drastically
in the  case of \ri: the  `field-to-particle dequantization'
and `quantization'  arrows on the diagram lead generally
 to different  QM in \ri  even  in the case of $T\otimes V_3$.
Therefore, the divergence is not caused  by  the particle
creation and annihilation processes. The unique
class of \ri,  when one can make consistent the  results of the two approaches
to QM is the Bianchi I type of which the spatially flat
Friedmann--Robertson--Walker (FRW) space-time is the isotropic subclass.
This is the main assertion in this my letter.

I should like to think it is not
an  accidental coincidence that  the spatially flat
FRW space-time is apparently realized apparently in   our Universe. At least,
astrophysical data testify  more and more convincingly, see, e.g.,
\cite{PRIM97},  to that
$$\Omega\ \ \defst\ \ \frac {\rho_b+\rho_d + \rho_\Lambda}{\rho_c}  =\ 1 $$
where $\rho_b$ is the average density of the ordinary (barionic )
matter in the Universe
$\rho_d$  is the density of the dark matter,  $\rho_c$ is the critical density
corresponding to the boundary between the spheric and hyperbolic
geometries of the space  and
$ \rho_\Lambda = \Lambda/(3{H_0})^2 $  is the effective
contribution of the cosmological constant $\Lambda$, $H_0 $being the Hubble
constant.   Of course, there should be a very deep  reason
for that the spatial curvature  in the Universe is
equal or very close to zero, the value which is
 critical between the continua of possible closed
and open FRW cosmologies.   One might say joking that  there exists
some  sort  of "quantropic  principle" which fixes  the geometry
so that internally consistent quantum theory could exist.

Consider the situation with QM in \ri and  especially in the
FRW space-times  in some detail on the basis of
\cite{TAG99, TAGI99}.

As concerns  traditional operator quantization of
a  Hamiltonian system,  the canonical commutation relations (CCR)
\beq
[\hpif,\ \hpjf]\ =\ 0, \quad [\hqif, \ \hqjf],\ =\ 0, \quad
[\hqif,\ \hpjf]\ = {\rm i} \hbar {\delta^{(i)}}_{(j)} \cdot\ {\bf\hat 1}
\label{ccr}
\nde
take a central place among its postulates. Here,  the basic operators
\hqi and \hpj, $i,\ j,\ k, ... = 1,\ 2,\ 3 $,  of position  and conjugate
momentum observables acting on a Hilbert space of states $\cal H$
correspond to the Darboux coordinates
$ \oqif, \ \opjf $ on the  phase space $  T^*\Sg_3 $,  the cotangent
bundle over a Cauchy hypersurface   $\Sg_3$.
(Closing  indices in the parentheses  denotes that the former
refer to the phase space.)
Thus,  $\Sg_3$ is the configurational space and provides \ri by a
1+3-foliation  (a frame of reference in the physical terms)
by a one--parametric system of Cauchy hypersurfaces $\Sg_3 \{s\} $
which are  normal geodesic translation of $\Sg_3\ \equiv\ \Sg_3 \{0\} $.
The latter  is assumed here   to be a  topologically
elementary manifold,  which means, in fact,   that only local
physical manifestations of curvature are taken into account in
the region where the normal geodesic congruence has no
focal points.

A  quantum-mechanical  operator $\hat f$ corresponding to a given
function (classical observable) $f(q,\ p)$  is supposed
to be constructed  by some procedure from the basic operators
$\hqif,\ \hpjf$
and the  function  itself  $f(q,\ p)$. \footnote{It should be
remarked here that,  from the pragmatic
physicist point of view of a pragmatic
physicist, to describe the quantum motion of a particle,
one  needs actually  only the Hamilton operator in addition
to \hqi and \hpi.  An alternative FT approach considered below
solves just this restricted problem.} There are infinitely many
procedures like that. For our  case of motion
of a structureless particle, these procedures are equivalent still
$\rif \sim E_{1,3}, \quad \Sg_3 \sim R_3 $ and
Cartesian coordinates $X^i$ on $\Sg_3$ and
 momenta $P_i$ conjugate to them are taken. If even one of these
three conditions is broken  the ambiguity  of the quantization map
$ f(q,\ p) \rightarrow \hat f$  becomes physically essential and manifests
itself in some way. In canonical quantization, this problem is
known as the problem of ordering of operator  products .
I leave  this  very important topic for a special discussion
elsewhere; some idea of the problem in
 canonical quantization and paths integration approaches  can be found
in  \cite{OLIV89, TAGI99}.\

The problem of product ordering  in  canonical quantization is combined
with another ambiguity, namely,   a dependence of $\hat f$  on choice
of coordinates \oqi even a rule  for each fixed ordering.
This means, for example, that the quantum Hamilton operator
determining the dynamics of the  system  under consideration
depends on the choice of classical observables  \oqi, \oqj. Thus,
along the `quntization' arrow
 in the diagram above, one comes to infinitely many QMs for
the same classical mechanics. Again,  in  $\Sg_3 \sim R_3$,
there exists a preferred  class $\oqif = X^i$ determined by the
isometry group of space translations.

I concentrated here on the  problems of the canonical operator
approach to quantization, but it is known  that such  popular
alternative as paths integration has an equivalent ambiguities,
see, e.g.,  \cite{OLIV89}.
Restrict now our consideration of QM in \ri to the general FRW space-time.
There exists
a natural 1+3-foliation which reduces
the metric to  the form, see, e.g., \cite {WEIN75}, Sec.14.2,
\beq
ds^2\ = \ c^2 dt^2\ - \ b^2 (t)\ \omega_{ij} (\xi; k)\ d\xi^i d\xi^j,
\label{ds}
\nde
where  $\omega_{ij}\ (\xi; k) $ is the metric tensor of a space section
$ t= const $  which is a 3-sphere, a 3-plane and
a 3-hypershere respectively for $ k= 1, \ k= 0 $ and $ k= -1 $.
The classical Hamilton function
$ H (q,\ p;\ t),\quad  \{q,\ p\} \in T^*\Sg_3\{t\} $
for a geodesic motion $\xi^i \ = \  \xi^i (t) $  is,
see, e.g. \cite{TAGI99}
\beq
 H (\xi,\ p;\ t)\ =\  mc^2 \left( 1\ - \
\frac{\omega^{ij} (\xi) \ p_i p_j }{b^2 (t)\ m^2 c^2} \right)^{1/2}
\label{hmc}
\nde

We see  that the coordinates $\xi^i $  have a
two-fold  purpose in the Hamilton formalism:

a) to provide $\Sg_3 \equiv \Sg_3 \{t\}$ by a manifold structure
(arithmetization):
$$
    \Sg_3\ \supset\ U\ \stackrel{\xi^i}{\longrightarrow}\ {\Bbb R}_3;
$$

b)  to be a classical observable of position of the particle
as if it were $ \oqif \equiv \xi^i $.

It is useful, though not necessary, to separate these purposes
using $\{\xi^i\}$  only for purpose a) ( that is to consider them as
an ordinary coordinate system on $\Sg_3$)  and introducing  general Darboux
coordinates $ \{\oqif, \opjf\} $ in the phase space
by the following map:
\beq
\xi^i   \longrightarrow \oqif (\xi), \quad
p_j  \longrightarrow  \opjf\ =\ K_{(j)}^l (\xi) p_l,
\quad    K_{(j)}^l \ptl_l \oqif\ = \ \delta_{(j)}^{(i)}.
\nde
Thus, the observables of spatial position  and momentum are defined
by values of  $\oqif (\xi)$  and $\opif (\xi)$  which are scalar fields
with respect to diffeomorphisms of $ \Sg_3 \{t\} $ for each fixed
value of $ t $. Now we can rewrite the Hamilton function  as
\beq
H (\oqif,\ \opjf;\ t)\   =\  mc^2 \left(1\ - \
\frac{\omega^{(kl)} (q) \ p_{(k)} p_{(l)} }{b^2 (t)\ m^2 c^2}\right )^{1/2}
\label{hmc2}
\nde
where $ \omega^{(kl)} (q)\
=\ \ptl_i q^{(k)}\ \omega^{ij} (\xi) \ptl_jq^{(l)} $   is
now a scalar with respect to the diffeomorphisms  of $\Sg_3$  and
a classical  observable as a function on the phase space.

Then, the corresponding basic  operators   can be represented
as differential operators acting
in ${\cal H} = L^2 (\Sg_3;\ {\Bbb C};\ b^3(t) \sqrt \omega\ d^3 \xi) $,
(that is, in the space of complex functions $\psi (\xi;\ t)$ which
are  square integrable  over $\Sg_3 (t)$ with the  natural measure
$d\sigma \defst b^3(t) \sqrt \omega\ d^3 \xi) $  and  have
the standard Born probabilistic interpretation):
\beq
\hqif(\xi) \ \defst \ \oqif(\xi)\cdot {\bf \hat 1},\quad
\hpjf(\xi)\   \defst\  -i\h\ \left(K_{(j)}^l (\xi) \tilde\nabla_l\
+ \ \frac12\ \tilde\nabla_l K_{(j)}^l (\xi)\right) ,            \label{qp}
\nde
 where $\tilde\nabla_l $ is the covariant derivative determined
by the metric tensor  $\omega_{ij} $. {\it Hence and throughout
the "hat" over characters  denotes  differential operators  with variable
coefficients, which act in
$L^2 (\Sg_3 (t);\ {\Bbb C};\ b^3 (t)  \sqrt \omega\ d^3 \xi)$ and
 contain only derivatives along $\Sg_3 (t)$ for fixed $t$}.
The  Hamilton operator corresponding to $ H (q,\ p ;\ t) $, eq.(\ref{hmc})
 is  obtained from the latter by substitution   the operators
\footnote{It is remarkable that now the metric tensor has become, in
a definite sense, quantized!}
 $\hqif(\xi) ,\  \hpjf (\xi)$ and
$ {\hat\omega}^{(ij)} (\xi)\ =\ \omega^{(ij)} ({\hat q}) =
\omega^{(ij)} (\xi) \cdot {\bf 1} $  {\it in some order}
instead of \oqi , \opj  and $\omega^{ij} (\xi)$ into  $H (q,\ p ;\ t) $
Then for any ordering one obtains
\beq
  H_0 \defst \frac1{2m}\ \omega^{(ij)}\ p_{(i)} p_{(j)} \quad
\stc{quantization}{\longrightarrow}\quad {\hat H_0} \ = \
 -\ \frac{\h^2}{2m} \Delta_\Sg \
+ \ V_q(\ptl q, \tilde\nabla\ptl q, ...; t ), \label{h0}
\nde
and, after a unitary transformation, one has \cite{TAGI99}
\beq
\hat H (\hat q,\ \hat p; \ t) \ = \   mc^2 \Biggl(
\sqrt{1\ +\ \frac{2 \hat H_0}{mc^2}}\ -\ 1 \Biggr),
\label{hmc3}
\nde
We see that  owing to  $V_q$
the Hamilton operator and, consequently,
the quantum dynamics  depends on the  choice of observables $q^{i} (x)$ !
Thus, even if we postulate a concrete rule of ordering (e.g.,
Weyl's one
is very popular),   we nevertheless  have an infinite  variety of QMs
instead of a single firmly established theory. However, in the spatially
flat  FRW space-time,  a preferred choice of \oqi exists: $\oqif = X^i$.
If it is done, then not only the quantum potential is
fixed as equal to zero, but also  the problem of ordering disappears
since  $\omega^{(ij)} = const $.

Now, let us go along to the  `field-to-particle' arrow on our diagram above
following paper \cite{TAG99}.
In this approach,  the  one-quasiparticle subspace $ \Phi^- $ of
Fock space ${\cal F}$ for the free quantum scalar field
\footnote{Below,  unlike  \cite{TAG99}, the operators
acting in ${\cal F}$ will be denoted by as $\check O$ to
be distinguished from  the QM-operators denoted as $\hat O$.}
$\check \varphi$ is constructed by  an  analogy with   the
standard quantum--mechanical  concept of localized particle,
see, e.g., \cite{JAU68} and  a generalization of the concept
to \ri in \cite{TAGI99}. This means that

a) $ \Phi^- $  is mapped
asymptotically in $c^{-2}$ onto $L^2 (\Sg;\ {\Bbb C};\ d\sigma) $,
see  Sec 4. in \cite{TAG99};
thus, $\psi (x)\ \in \ L^2 (\Sg;\ {\Bbb C};\ d\sigma) $ can be considered
as the probability amplitude to find the particle at the point $x \in \Sg$
and further results are comparable with those of the canonical
quantization;

b) operators of basic  one-particle observables acting  in this
$L^2 (\Sg;\ {\Bbb C};\ d\sigma) $  as differential operators are
generated by the  corresponding  field-theoretical (FT) operators
acting in   ${\cal F}$;

c) the Hamilton operator $H (x)$ is determined by
asymptotic transformation of the field equation for $\vp$ to
the \Sche.
\noindent
The one-quasiparticle  subspace $ \Phi^- $ thus determined may be
considered as one-particle subspace of ${\cal F}$.

The field equation  is the well-known generalization of the
Klein--Gordon--Fock equation to \ri in  the general non-minimal form
 \begin{eqnarray}
\Box\varphi + \zeta\, R(x)\, \varphi  + \left(\frac{mc}{\hbar}
\right)^2  \varphi = 0,  \quad x\in \rif \label{r} \\
\Box \stackrel{def}{=}  g^{\alpha\beta}\nabla_\alpha \nabla_\beta,
\qquad  \alpha,\beta,... = 0,1,2,3,   \nonumber
\end{eqnarray}
$R (x)$ is the scalar curvature of \ri ($R(x) = R(t)$ in the
FRW space-time.),  and  $\zeta$ is   a free parameter .
This equation generates asymptotically in $c^{-2}$   the following \Sche  
in the FRW space-time
for $\psi(x) \equiv \psi(\xi;\ t) \ \in\
L^2 (\Sg_3 (t) ;\ {\Bbb C};\ b^3 (t) {\sqrt \omega (t)}  d^3\xi)$:
\beq
   {\rm i}\h{\cal T}\ \psi  \ =\ {\hat H}_N (\xi;\ t)\ \psi;
 \qquad  {\cal T} \ \defst\   \frac\ptl{\ptl t}\  +
\frac32 \frac{\ptl b(t)}{\ptl t}.
\nde
Here ${\hat H}_N$  is the Hamilton operator which is an asymptotic
expansion starting with $\hat H_0$, eq.(\ref{h0}), in which it is taken
$V_q = -(\h^2/2m)\ \zeta\ R $. For $N\rightarrow\infty$, this
expansion  can formally be partially summed  and represented
through ${\hat H}$, eq.(\ref{hmc3}), as follows:
\beq
{\hat H}_\infty \ = \ {\hat H}\ +\
\sum\limits_{n=1}^\infty \ \frac{{\hat{\tilde h}}_n (\xi;\ t)}{(2mc^2)^n}
\nde
and the operators  ${\hat{\tilde h}}_n (\xi;\ t)$ are such that they
vanish if $ [{\cal T},\ H_0]\ = \ 0 $.
Note that ${\cal T}$ is not an operator in
 $L^2 (\Sg_3 (t);\ {\Bbb C};\ b^3 (t)\ \sqrt{ \omega}\ d^3 \xi)$.
Contrary to the canonical approach,
 the Hamilton operator is   a scalar with respect to diffeomorphisms
of $\Sg_3 (t)$ and depends only on choice of the parameter $\zeta$.
The latter, in turn, does not affect the main conclusions of the present
letter.

The operators of  basic observables of position  and momentum are now 
determined
 asymptotically by  some QFT-operators (operators  acting in $\cal F$).
As concerns the operator  ${\hat p}_K (x) $ of projection of
momentum ${\hat p}_K (x) $ on any  given vector field $ K^\alpha (x)$,
not necessarily directed
along $\Sg_3 (t)$, it is natural to determine it through the
corresponding  QFT--operator  for the quantized scalar field
$\check \varphi (x)$  and given  $\Sg$:
\begin{equation}
{\check{\cal P}}_K \{\check\varphi ; \, \Sg\} \ = \  :\int_\Sg
d\sigma^\alpha \ K^\beta T_{\alpha\beta} (\check\varphi):, \label{pk}
\end{equation}
where the colons denote the normal product of the creation and
annihilation  operators in ${\cal F}$
and $T_{\alpha\beta}$ is the metric energy--momentum tensor for
$\varphi (x)$  \cite{CHT68}; in  the FRW case, it is natural to put
$\Sg \sim \Sg_3 (t) $.  Then, for
$ {K_{(i)}}^\alpha (x)  \defst {K_{(i)}}^j \delta_j^{\alpha}$ the
matrix elements of
${\check {\cal P}}_{K_{(i)}} \{\hat\varphi ;\, \Sg_3 (t)\}$  in $\Phi^-$,
being expressed as matrix elements of an operator
in $L^2 (\Sg_3(t);\ {\Bbb C};\ b^3(t)\ \sqrt{\omega}\ d^3 \xi)$,
give  the field--theoretically determined quantum--mechanical operator
of projection  momentum of a localizable configuration of the quantum field
${\check\varphi} (x)$. It  can be calculated as  asymptotic expansion
${\hat p}_{(i),N} (\xi;\ t)$
from the general and generally covariant expression (53) in \cite{TAG99}.
 Again, ${\hat p}_{(i),0} (\xi;\ t)= {\hat p}_{(i)} (\xi;\ t)$  being
defined in eq.(\ref{qp}). In Section 5 of \cite{TAG99}  it is
 shown also that the triple ${\hat p}_{(i),N} (\xi;\ t)$ is commutative
for any $N$  if $K_{(i)}^j $ are  commutative Killing vector fields on
$\Sg_3 (t)$. {\it Just in this case \ri is of the Bianchi I type.}
Otherwise, the asymptotic terms deform the  first relation in CCR,
eq.(\ref{ccr}).

It remains now to consider, in the same way,   relations
involving spatial position observables.
Operators   of spatial position of a particle in QFT have been considered in
\cite{POLU73}, \cite{DUR76}, but only  in terms of the  Cartesian
coordinates in the Minkowski space-time. However,  since the QFT--prototype
$\check {\cal P}_{K_{(i)}} \{\hat\varphi ;\ \Sg)\} $
for $p_{(i)} (\xi;\ t)$
exists one may expect that there also exist   an analogous
generally covariant QFT--operators $\check Q^{(i)}$.  By
analogy with  $\check {\cal P}_{K_{(i)}}$,  they
should be integrals
over $\Sg$ of a sesquilinear form of $\check \varphi$ and  linear
functionals of  arbitrary scalar functions  $q^{(i)} (x) $ which  satisfy
the conditions
$ \ptl^\alpha \Sg(x)  \ptl_\alpha q^{(i)} (x)\ =\ 0, \quad
{\rm rank}|| \ptl_\alpha q^{(i)} (x)||\ =\ 3 $ and thus
determine a point on each $\Sg_3 (t)$. It is natural, in the FRW case,
to adjust their values with $q^(i) (\xi) $  introduced
in the canonical approach.

It appears that
for a given triple $q^{(i)} (x)$  there is a unique triple
of QFT--operators
\beq
\check{\cal Q}^{(i)}\{\check\varphi;\, \Sg_3 (t)\} =
 \int_\Sg \ d^3 \xi b^3(t) \sqrt{\omega}\ q^{(i)}(\xi)\ {\check N} (\xi;\ t)
\label{Q2}
\nde
where  ${\check N} (\xi;\ t)$ is the QFT--operator of quasiparticle
density. (Here I rewrite the generally covariant eq.(19)
from \cite{TAG99} for the particular case of the  FRW metric (\ref{ds}).

In the same way, as for  $\check {\cal P}_K \{\hat\varphi ; \Sg)\} $,
one comes to the general formula (59) in \cite{TAG99} which
is in fact  an  asymptotic expansion  of the form
\beq
{\hat q}^{(i)}_N (\xi;\ t) = {\hat q}^{(i)} (\xi;\ t)
\ +\  \sum\limits_{n=2}^N \ 
\frac{{\hat {\tilde q}}^{(i)}_n (\xi;\ t)}{(2mc^2)^n}
\ + \ O (c^{-2(N+1)}). \label{qn}
\nde
{\it Note that the corrections to ${\hat q}^{(i)}$ start with a
term of  order  $O (c^{-4})$}.

It is easy to see  from consideration of
${\hat {\tilde q}}^{(i)}_2 (\xi;\ t)$,
that operators ${\hat q}^{(i)}_2 (\xi;\ t)$ do not commute
except the case when   $ q^{(i)} (\xi;\ t) \equiv \xi^i \equiv X^i,
 \quad X^i $ being Cartesian coordinates in a Bianchi I  space-time.
It remains to show that  asymptotic operators
$\hat X^i_N$ mutually  commute  for $ N \rightarrow \infty$ too.
It is not trivial because, owing to a time dependence of the metric,
eq.(59) in \cite{TAG99} and its  expression in the form (\ref{qn}) give
an infinite series for $\hat X^i_\infty$.
 I restrict the consideration by the FRW case; generalization to
the Bianchi I type is straightforward.  Then, the following
proposition can be easily proved.\\
\noindent
{\it Proposition.} \\
In the spatially flat  FRW space-time, a unitary operator $\hat U$ exists  
such that
\beq
     \hat U\ {\hat X}^i_{\infty}, \  {\hat U}^\dagger 
= X^i \cdot {\bf\hat 1} .
\label{X}
\nde
{\it Proof.}
At first, owing to the  translational invariance
 $ \hat U$\ and\  ${\hat X}^i$ have correspondingly the forms
\beq
 \hat U \ = \ \sum\limits_{n=0}^\infty \ u_n (t) \Delta^n, \quad
{\hat X}^i_{\infty}\ = \ X^i \cdot {\bf\hat 1}\ +\
 \sum\limits_{n=0}^\infty \ x_n (t) \Delta^n\ \frac\ptl{\ptl X_i} \label{X}
\nde
where $\Delta$ is the Euclidean Laplace operator. Therefore eq.(\ref{X})
can be represented in the form
\beq
[\hat U,\ X^i\ {\bf\hat 1}]\ 
= \ \hat U \ {\cal X}(t, \Delta) \frac\ptl{\ptl X_i}
\nde
which is equivalent to the  equation
$$  \frac\ptl{\ptl s } U(t, s)\ = \ \frac{\rm i}2
\sum\limits_{n=0}^\infty \ x_n (t) s^n \ U(t,s);\quad
U(t, s) \defst \sum\limits_{n=0}^\infty \ u_n (t) s^n
$$
which determines  the coefficients $ u_n (t) $ and always has a solution
such that the condition of unitarity
$\hat U\ {\hat U}^\dagger = {\bf\hat 1}$
is fulfilled. \quad  Q.E.D.

Thus, the  algebraic structures of basic  observables
of spatial position and momentum in the canonical and field-theoretical
approaches to QM in \ri
are in accordance   only when \ri is of the Bianchi I
type and the Cartesian coordinates are taken  as the
classical position observables.
Has this  fact any concern
to the  observed spatial flatness of the  Universe, or this
is only an accidental coincidence, seems to be a question of fundamental
interest. To study   consequences of
non-commutativity  of coordinates  seem not less interesting \\

{\bf Acknowledgement}\\

The author is deeply grateful to the Russian Foundation for
Basic Research which supported this work  by Grant No 00-01-00871.

\end{document}